\documentclass[10pt,a4paper]{article}
\usepackage{amsmath,amssymb,amsfonts}
\usepackage{graphicx}
\usepackage{hyperref}
\usepackage[margin=0.75in]{geometry}
\usepackage{float}
\usepackage{bm}
\usepackage{booktabs}
\usepackage{tabularx}
\usepackage{subcaption}

\usepackage{setspace}
\usepackage{titlesec}
\titlespacing*{\section}{0pt}{8pt}{4pt}
\titlespacing*{\subsection}{0pt}{6pt}{3pt}

\AtBeginDocument{
  \setlength{\abovedisplayskip}{4pt}
  \setlength{\belowdisplayskip}{4pt}
  \setlength{\abovedisplayshortskip}{2pt}
  \setlength{\belowdisplayshortskip}{2pt}
}

\let\oldbibliography\thebibliography
\renewcommand{\thebibliography}[1]{%
  \oldbibliography{#1}%
  \setlength{\itemsep}{0pt}%
  \setlength{\parskip}{0pt}%
}

\title{Conservation Constraints and Distributed Advective Memory in a
Reduced Model of Atlantic Overturning Hysteresis}

\author{
Sandy Hardian Susanto Herho$^{1,2}$,
Iwan Pramesti Anwar$^{3}$,
Mutiara Rachmat Putri$^{3}$,\\
Rusmawan Suwarman$^{4}$,
Deny Juanda Puradimaja$^{1}$, and
Dasapta Erwin Irawan$^{1,*}$
}

\date{}

\begin{document}
\maketitle

\begin{center}
\small
$^{1}$Applied Geology Research Group, Bandung Institute of Technology,
Bandung, West Java 40132, Indonesia\\
$^{2}$Center for Agrarian Studies, Bandung Institute of Technology,
Bandung, West Java 40132, Indonesia\\
$^{3}$Applied and Environmental Oceanography Research Group, Bandung
Institute of Technology, Bandung, West Java 40132, Indonesia\\
$^{4}$Atmospheric Science Research Group, Bandung Institute of Technology,
Bandung, West Java 40132, Indonesia\\
$^{*}$Correspondence: dasaptaerwin@itb.ac.id
\end{center}

\begin{abstract}
\noindent
Interbasin exchange through the Indo-Pacific gateway supplies salt to the
Atlantic and is widely invoked as a control on the stability of the Atlantic
overturning circulation. We ask whether that control can act on
the equilibrium structure of a conceptual thermohaline model. A closed
five-box model with an exact salt invariant is constructed, comprising North
Atlantic, upper-limb, Indian, Pacific, and deep reservoirs, with the return
flow split between a warm route through the Indian reservoir and a cold
route, together with an Indonesian Throughflow branch and an Agulhas
retroflection. Adding the steady-state budgets of the gateway
reservoirs shows that every internal exchange cancels, so the salt they
export to the Atlantic is fixed by the net Atlantic freshwater export alone.
This holds independently of the warm-route fraction, the throughflow, the
retroflection, and how the export is apportioned among gateway reservoirs;
across a parameter sweep the largest departure is of order ten to the minus
eleven. The gateway therefore enters as a purely additive
forcing and cannot renormalize the salt-advection feedback. Replacing the
discrete transit lag by a gamma memory kernel leaves the equilibria
unchanged but yields a closed-form threshold for oscillatory instability
depending only on kernel shape. Broad memory is strongly stabilizing, and a
discrete lag is the least stable member of the family. Because the
instantaneous feedback vanishes at the fold, oscillatory instability always
precedes the saddle-node, over an interval widening more than tenfold as
memory sharpens. Gateways therefore appear to act on transient rather than
equilibrium dynamics.
\end{abstract}


\section{Introduction}
\label{sec:intro}

The Atlantic meridional overturning circulation (AMOC) transports heat and
salt northward in the Atlantic basin and is among the most closely examined
candidate tipping elements of the climate system \cite{Lenton2008,
ArmstrongMcKay2022, Weijer2019, Dijkstra2026}. Its dynamical interest originates in the
demonstration that a two-reservoir system exchanging heat and salt admits
two stable regimes of flow, with the transition between them governed by a
positive feedback in which the circulation advects the salt that sustains
its own density contrast \cite{Stommel1961}. A distinct destabilization route, in which convective
overturning switches on and off, has also been identified
\cite{Welander1982}, and the two mechanisms have since been shown to
organize the bifurcation structure of models throughout the hierarchy
\cite{WeijerDijkstra2001, DijkstraWeijer2005}. In the limit of fast thermal
relaxation the two-reservoir system reduces to a single scalar equation
whose quartic potential has two wells, corresponding to the active and
collapsed circulations \cite{Cessi1994}. The resulting hysteresis in freshwater
forcing has since been reproduced across a hierarchy of models, from box
formulations through Earth system models of intermediate complexity
(EMICs) \cite{Rahmstorf2005} to coupled general circulation models (GCMs)
\cite{vanWesten2024}.

Whether the present-day ocean lies inside or outside the bistable window
remains contested. Of eleven EMICs subjected to a standardized freshwater
perturbation, seven placed present-day climate in the bistable regime and
four in the monostable regime, with hysteresis widths spanning $0.2$ to
$0.5$ sverdrups (Sv) \cite{Rahmstorf2005}, where
$1~\mathrm{Sv} \equiv 10^{6}~\mathrm{m^{3}\,s^{-1}}$.
Statistical early-warning indicators applied to observational proxies have
been read as evidence of destabilization \cite{Boers2021}, and data-driven
estimators have been used to assign a time of tipping
\cite{Ditlevsen2023}, while a
physics-based indicator derived from the overturning freshwater transport at
the Atlantic southern boundary has been argued to place the present-day
circulation on a tipping trajectory \cite{vanWesten2024}. Against this,
Southern Ocean wind-driven upwelling has been found to sustain a weakened
circulation across 34 models under extreme forcing, so that complete
collapse appears unlikely within the present century \cite{Baker2025}. This divergence motivates continued scrutiny of the mechanisms that
conceptual models attribute to bistability.

One such mechanism is interbasin exchange. The upper-layer return flow
compensating North Atlantic Deep Water (NADW) formation has been argued to be
routed largely through the warm thermocline, linking the Atlantic to the
Indian Ocean \cite{Gordon1986}. The Agulhas system delivers salty Indian Ocean
thermocline water into the South Atlantic \cite{Beal2011}, and the
Indonesian Throughflow (ITF) sets the salinity of the Indian Ocean reservoir
that the leakage subsequently inherits \cite{LeBars2013}. Because the Atlantic
salinity surplus is maintained in part by this import \cite{Weijer2019}, it
is natural to suppose that the partition of the return flow between the warm
and cold routes modulates the strength of the salt-advection feedback, and
with it the existence and width of the hysteresis loop. The observational
record supports that supposition only indirectly. Multi-method causality
analysis of monthly transport records spanning 1984 to 2023 recovers several
significant pathways from ITF components to Agulhas variability at lags of
zero to eighteen months, but no detectable direct connection from either the
ITF or the Agulhas system to the AMOC at 26$^{\circ}$N \cite{Herho2026}.
That null result admits more than one reading, among them the limited
meridional coherence of the Atlantic overturning and the brevity of the
observational window relative to gateway transit times, and it does not on
its own settle whether the return-flow partition acts on the equilibrium
structure. Whether that
supposition survives a budget in which salt is conserved exactly has not, to
our knowledge, been checked.

A second line of inquiry concerns the finite time required for salinity
anomalies to traverse the circulation. Conceptual AMOC models are
conventionally formulated as ordinary differential equations (ODEs) in which
the feedback acts instantaneously \cite{Stommel1961, Cessi1994, Wood2019,
Alkhayuon2019}. Advective transit, however, spreads arrival times over a
distribution rather than concentrating them at a single lag, because
anomalies follow many paths. Directional information-flow analysis of
throughflow transport records is consistent with that picture: the lag at
which transfer entropy from a climate index into the throughflow is
maximized varies by transport component, from four to nine months, rather
than collapsing onto a common value \cite{Herho2025}. In other fields where dwell times are
explicitly distributed, replacing a discrete lag by a gamma memory kernel is
known to alter stability thresholds, and the linear chain trick provides an
exact finite-dimensional representation of such kernels
\cite{Hurtado2019, Ando2020}. The consequences of this substitution for
thermohaline bistability do not appear to have been examined.

These two questions turn out to be connected. A box model in which the total
salt is an exact invariant permits the gateway budget to be summed in closed
form, and doing so settles the first question at once, in the negative;
having closed the equilibrium channel, the memory kernel becomes the only
route left by which the gateway can act, which is what makes the second
question worth pursuing rather than merely tractable. We accordingly develop
both within a single conservative five-box formulation, deriving the
equilibrium structure, the Hopf threshold under a gamma kernel, and the
forcing interval separating that threshold from the fold. Nothing here rests
on GCM output or on observational or reanalysis data; every quantity
reported is obtained from the model equations and, wherever a closed form
exists, checked against it.
\section{Methods}
\label{sec:methods}

\subsection{Model description}
\label{subsec:model}

Consider first an ocean idealized as two well-mixed reservoirs of equal
volume $V$, a low-latitude reservoir labeled $1$ and a high-latitude
reservoir labeled $2$, connected so that a volume flux $q$ passes from one
to the other and an equal flux returns \cite{Stommel1961}. Let $T_{j}$ and $S_{j}$ denote the
temperature and salinity of reservoir $j$. Conservation of heat and salt
gives
\begin{equation}
    V\frac{\mathrm{d}T_{1}}{\mathrm{d}t}
      = q\left(T_{2}-T_{1}\right)
      + \frac{V}{t_{r}}\left(T_{1}^{\ast}-T_{1}\right),
    \qquad
    V\frac{\mathrm{d}T_{2}}{\mathrm{d}t}
      = q\left(T_{1}-T_{2}\right)
      + \frac{V}{t_{r}}\left(T_{2}^{\ast}-T_{2}\right),
    \label{eq:heat_boxes}
\end{equation}
where $T_{j}^{\ast}$ is the atmospheric temperature toward which reservoir
$j$ is restored and $t_{r}$ is the restoring timescale. Salinity is not
restored but is forced by a net freshwater flux $F$, which for an
evaporative low latitude and a precipitative high latitude removes salt from
reservoir $2$ and adds it to reservoir $1$ at a rate $F S_{0}/H$ per unit
depth, with $S_{0}$ a reference salinity and $H$ a mixed-layer depth. This
virtual salt flux representation, in which freshwater exchange is converted
to an equivalent salt flux at fixed volume, is the standard treatment for
conceptual models of this class \cite{Rahmstorf1996}:
\begin{equation}
    V\frac{\mathrm{d}S_{1}}{\mathrm{d}t}
      = q\left(S_{2}-S_{1}\right) + \frac{V F S_{0}}{2H},
    \qquad
    V\frac{\mathrm{d}S_{2}}{\mathrm{d}t}
      = q\left(S_{1}-S_{2}\right) - \frac{V F S_{0}}{2H}.
    \label{eq:salt_boxes}
\end{equation}
Only the interreservoir contrasts drive the circulation, so we subtract
within each pair and introduce
\begin{equation}
    T \equiv T_{1}-T_{2},
    \qquad
    S \equiv S_{1}-S_{2},
    \qquad
    \theta \equiv T_{1}^{\ast}-T_{2}^{\ast},
    \qquad
    \mathcal{Q} \equiv \frac{2q}{V},
    \label{eq:contrasts}
\end{equation}
where $\mathcal{Q}$ has dimensions of inverse time and is the rate at which
a contrast is destroyed by exchange. Equations~(\ref{eq:heat_boxes}) and
(\ref{eq:salt_boxes}) then collapse to the pair
\begin{equation}
    \frac{\mathrm{d}T}{\mathrm{d}t}
      = -\frac{T-\theta}{t_{r}} - \mathcal{Q}\,T,
    \qquad
    \frac{\mathrm{d}S}{\mathrm{d}t}
      = \frac{F S_{0}}{H} - \mathcal{Q}\,S.
    \label{eq:TS}
\end{equation}

Closing the system requires an expression for $\mathcal{Q}$ in terms of the
state. Adopting a linear equation of state \cite{Stommel1961, Cessi1994},
\begin{equation}
    \rho_{j} = \rho_{0}\left(1 - \alpha T_{j} + \beta S_{j}\right),
    \label{eq:eos}
\end{equation}
with $\rho_{0}$ a reference density, $\alpha$ the thermal expansion
coefficient, and $\beta$ the haline contraction coefficient, the density
contrast driving the exchange is
$\rho_{2}-\rho_{1} = \rho_{0}\left(\alpha T - \beta S\right)$. Two physical
requirements constrain how $\mathcal{Q}$ depends on this contrast. It must
remain strictly positive when the contrast vanishes, because molecular and
eddy diffusion continue to exchange properties in the absence of an
overturning flow; and it must be invariant under reversal of the sign of the
contrast, because a reversed circulation destroys a contrast just as
effectively as a direct one. The first requirement introduces a constant
diffusive floor $1/t_{d}$, and the second requires the flow-dependent
contribution to be an even function of $\alpha T - \beta S$. Retaining the
leading even term gives the quadratic closure \cite{Cessi1994},
\begin{equation}
    \mathcal{Q}(T,S) \equiv \frac{1}{t_{d}}
      + \frac{q_{a}}{V}\left(\alpha T - \beta S\right)^{2},
    \label{eq:closure}
\end{equation}
where $q_{a}$ is a transport coefficient and $t_{d}$ the diffusive
timescale. This is a smooth surrogate for the piecewise expression
$\mathcal{Q} \propto |\alpha T - \beta S|$ obtained when the exchange rate is
taken proportional to the magnitude of a density-driven flow
\cite{Stommel1961}.

Ocean thermal restoring is fast relative to the diffusive exchange, so the
ratio $\varepsilon = t_{r}/t_{d}$ is small; being the quotient of two
timescales, $\varepsilon$ is dimensionless, which is what permits it to be
used as an asymptotic ordering parameter in the first place. Writing
Eq.~(\ref{eq:TS}) in units of $t_{d}$ makes the temperature equation
$\varepsilon\,\mathrm{d}T/\mathrm{d}\tau = -(T-\theta)
- \varepsilon\,\mathcal{Q}t_{d}T$, so that at leading order in $\varepsilon$
the temperature contrast is slaved to the atmospheric contrast,
$T = \theta + \mathcal{O}(\varepsilon)$. Substituting $T \approx \theta$ into
Eq.~(\ref{eq:closure}) and introducing the non-dimensional salinity
contrast, time, forcing, and salt-advection gain
\begin{equation}
    x \equiv \frac{\beta S}{\alpha\theta}, \qquad
    \tau \equiv \frac{t}{t_{d}}, \qquad
    p \equiv \frac{\beta S_{0}t_{d}}{H\alpha\theta}\,F, \qquad
    m \equiv \frac{q_{a}t_{d}\left(\alpha\theta\right)^{2}}{V},
    \label{eq:nondim}
\end{equation}
gives $\alpha T - \beta S = \alpha\theta\,(1-x)$ and therefore
$\mathcal{Q}t_{d} = 1 + m(1-x)^{2}$. Each of these four groups is
dimensionless by construction, and for a distinct reason worth stating. The
contrast $x$ is the ratio of the haline to the thermal contribution to the
density anomaly in Eq.~(\ref{eq:eos}); both contributions carry the
dimensions of a relative density, so their quotient is a pure number, and
$x = 1$ marks exact compensation, at which the circulation is driven by
diffusion alone. The time $\tau$ is measured in units of the diffusive
timescale $t_{d}$, so that a unit interval corresponds to one exchange time
rather than to an arbitrary calendar interval. The gain $m$ is the ratio of
the advective to the diffusive contribution to $\mathcal{Q}$ evaluated at
zero salinity contrast, and therefore measures the strength of the
salt-advection feedback relative to the diffusive floor that opposes it.
The forcing $p$ is the freshwater flux expressed as the salinity contrast it
would generate over one diffusive timescale, again scaled by
$\alpha\theta/\beta$. Working in these variables is not merely
cosmetic: the ten dimensional quantities entering
Eqs.~(\ref{eq:heat_boxes})--(\ref{eq:closure}) collapse to the two groups
$m$ and $p$, so the entire bifurcation structure can be mapped in a plane
rather than explored in a ten-dimensional parameter space, and results
obtained at one set of physical scales transfer to any other set sharing the
same $m$ and $p$. The salinity equation in
Eq.~(\ref{eq:TS}) becomes the scalar evolution law
\begin{equation}
    \frac{\mathrm{d}x}{\mathrm{d}\tau}
      = f(x;p,m) \equiv p - x\left[1 + m(1-x)^{2}\right].
    \label{eq:scalar}
\end{equation}
The bracket is the non-dimensional exchange rate, which is largest when
$x = 0$ and smallest when $x = 1$, so the active circulation corresponds to
small $x$ and the collapsed circulation to $x$ near unity. The salt-advection
feedback is visible directly: an increase in $x$ reduces the bracket, which
reduces the rate at which the forcing-induced contrast is destroyed, which
increases $x$ further.

Equilibria of Eq.~(\ref{eq:scalar}) satisfy the cubic
\begin{equation}
    m x^{3} - 2m x^{2} + (1+m)x - p = 0,
    \label{eq:cubic}
\end{equation}
so that at most three real states coexist at a given forcing. Writing
subscripts for partial derivatives,
\begin{equation}
    f_{x} = -1 + m(1-x)(3x-1),
    \qquad
    f_{xx} = m(4-6x),
    \qquad
    f_{p} = 1,
    \label{eq:derivs}
\end{equation}
and a fold requires $f = f_{x} = 0$. The second condition rearranges to
$3x^{2}-4x+1+1/m = 0$, whose roots are
\begin{equation}
    x_{\pm} = \frac{2 \pm \sqrt{1-3/m}}{3},
    \label{eq:folds}
\end{equation}
real if and only if $m \geq 3$. The corresponding forcings follow from
$f = 0$ as $p_{\pm} = x_{\pm}\left[1+m(1-x_{\pm})^{2}\right]$, and the
bistable width is $\Delta p = |p_{+}-p_{-}|$. Imposing the further
degeneracy $f_{xx} = 0$ gives $x = 2/3$, which returned to the fold
condition yields $m(1/3)(1) = 1$ and hence
\begin{equation}
    \left(m_{c},\,x_{c},\,p_{c}\right)
      = \left(3,\ \tfrac{2}{3},\ \tfrac{8}{9}\right).
    \label{eq:cusp}
\end{equation}
Equations~(\ref{eq:folds}) and (\ref{eq:cusp}) are exact and serve as
references against which the numerical apparatus is checked.

\begin{figure}[H]
    \centering
    \includegraphics[width=\linewidth]{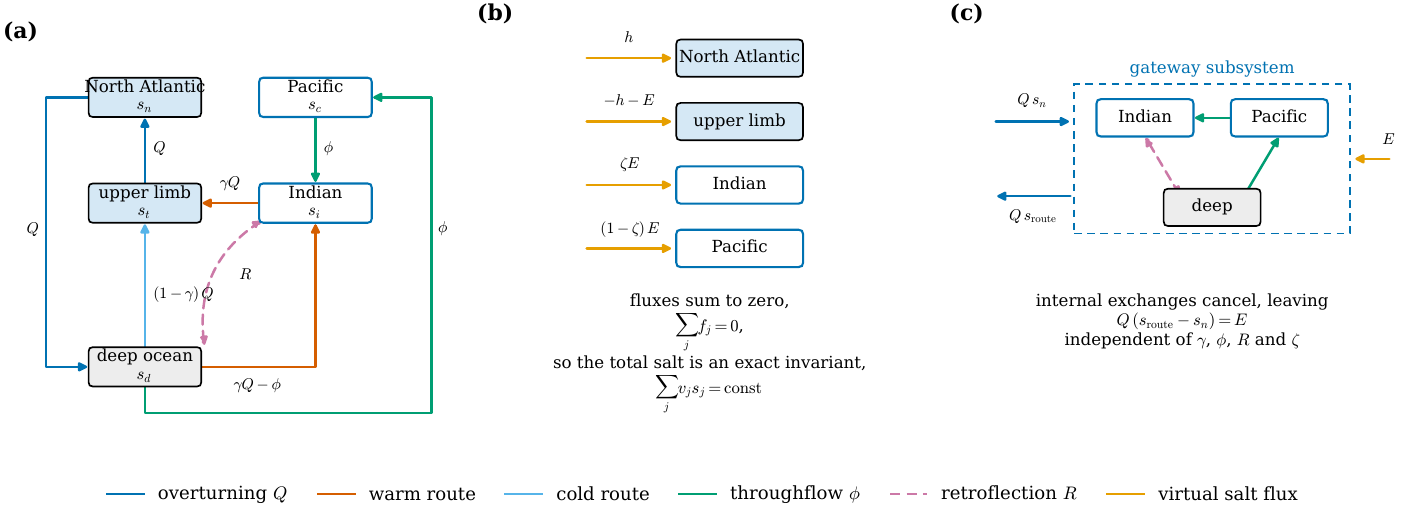}
    \caption{The closed five-box gateway model. (a) Circulation. The
    overturning $Q$ closes the loop $t \to n \to d$, with the return
    partitioned between a warm route through the Indian box, of magnitude
    $\gamma Q$, and a cold route of magnitude $(1-\gamma)Q$. The ITF branch
    $\phi$ carries deep water to the Pacific and Pacific water to the
    Indian box; the Agulhas retroflection $R$ is a bidirectional
    exchange between the Indian box and the deep ocean that bypasses the
    Atlantic. (b) Virtual salt fluxes, which sum to zero so that the
    volume-weighted total salt is an exact invariant. (c) The gateway
    subsystem, comprising the Indian, Pacific, and deep boxes, exchanges
    with the Atlantic only through $Q$ and receives net forcing $E$.}
    \label{fig:schematic}
\end{figure}

Interbasin exchange is introduced by extending the two-reservoir system to
five reservoirs while preserving exact salt conservation. Multi-reservoir
extensions of this kind have a long lineage, beginning with the
interhemispheric three-box configuration \cite{Rooth1982} and its later
elaborations \cite{Scott1999, Wood2019}; what distinguishes the present
construction is the requirement that the global salt budget close
identically, which is what makes the gateway argument below possible. The reservoirs are
the North Atlantic ($n$), the tropical and South Atlantic upper limb ($t$),
the Indian Ocean thermocline ($i$), the Pacific warm pool ($c$), and the
deep ocean ($d$), with non-dimensional volumes $v_{j}$ and
density-equivalent salinities $s_{j} \equiv \beta S_{j}/(\alpha\theta)$ for
$j \in \{n,t,i,c,d\}$. The volumes are dimensionless because they are
expressed as multiples of a common reference volume, so only their ratios
enter; the salinities are dimensionless because they are scaled exactly as
$x$ was above, which lets salinity and temperature contributions to density
be compared directly. All transports, namely $Q$, $\phi$, and $R$, are
likewise measured in units of the inverse diffusive timescale, so that
$Q = 1$ denotes a purely diffusive exchange. The partitions $\gamma$ and
$\zeta$ are dimensionless as fractions of a whole, each lying in $[0,1]$ by
definition. The topology appears in Fig.~\ref{fig:schematic}.

The overturning $Q$ is retained in the form obtained above,
\begin{equation}
    Q(x) \equiv 1 + \mu^{2}(1-x)^{2},
    \qquad
    x \equiv s_{t}-s_{n},
    \label{eq:Q}
\end{equation}
with $\mu^{2}$ the salt-advection gain of the extended system. The
circulation closes the loop $t \to n \to d$ and returns to the upper limb
along two routes: a fraction $\gamma \in [0,1]$, the warm-route fraction,
passes through the Indian box, and the remainder returns directly. Of the
flux entering the Indian box, a portion $\phi$ arrives by way of the
Pacific, representing the ITF, and the remainder arrives from the deep
ocean. The ITF is the only low-latitude interocean passage in the modern
configuration, and sustained mooring records place its Makassar Strait
component in the range of order $10$~Sv with strong monsoonal and
interannual modulation \cite{Susanto2012, Sprintall2014}, while geostrophic
transport estimates across the section between Indonesia and Australia
indicate a strengthening tendency over recent decades \cite{Herho2026}; the
present formulation retains only its role as a conduit of Pacific water into
the Indian reservoir. Its variability is concentrated at seasonal and
interannual periods \cite{Herho2025}, which are short compared with the
diffusive timescale, so $\phi$ is held fixed throughout. Requiring that the volume entering each reservoir
equal the volume leaving it fixes every branch:
\begin{equation}
    d \to i:\ \gamma Q-\phi,
    \quad
    d \to c:\ \phi,
    \quad
    c \to i:\ \phi,
    \quad
    i \to t:\ \gamma Q,
    \quad
    d \to t:\ (1-\gamma)Q,
    \quad
    t \to n \to d:\ Q,
    \label{eq:volume}
\end{equation}
which is consistent because the deep reservoir exports
$(\gamma Q-\phi)+\phi+(1-\gamma)Q = Q$ and imports $Q$. Superposed on this
advective skeleton is the Agulhas retroflection, a bidirectional exchange of
strength $R$ between the Indian box and the deep ocean which transports
equal volumes in each direction and therefore does not enter
Eq.~(\ref{eq:volume}). Assembling the salt budget of each reservoir as the
sum of advective imports minus advective exports, plus the retroflective
exchange, minus a virtual salt flux $f_{j}$, gives
\begin{align}
    v_{n}\dot{s}_{n} &= Q\left(s_{t}-s_{n}\right) - f_{n},
    \label{eq:box_n}\\
    v_{t}\dot{s}_{t} &= Q\left[\gamma s_{i}
      + (1-\gamma)s_{d} - s_{t}\right] - f_{t},
    \label{eq:box_t}\\
    v_{i}\dot{s}_{i} &= \left(\gamma Q-\phi\right)\left(s_{d}-s_{i}\right)
      + \phi\left(s_{c}-s_{i}\right)
      + R\left(s_{d}-s_{i}\right) - f_{i},
    \label{eq:box_i}\\
    v_{c}\dot{s}_{c} &= \phi\left(s_{d}-s_{c}\right) - f_{c},
    \label{eq:box_c}\\
    v_{d}\dot{s}_{d} &= Q\left(s_{n}-s_{d}\right)
      + R\left(s_{i}-s_{d}\right) - f_{d},
    \label{eq:box_d}
\end{align}
where an overdot denotes $\mathrm{d}/\mathrm{d}\tau$. Each advective term
appears as a relaxation of the receiving reservoir toward the salinity of
its source, which is the correct form when the outflow carries the
reservoir's own salinity and the inflow carries that of its origin.

The virtual salt fluxes encode two distinct processes. A hosing amplitude
$h$ freshens the North Atlantic and salinifies the upper limb, representing
an anomalous meridional redistribution of freshwater within the Atlantic.
Separately, the Atlantic is a net evaporative basin, exporting a freshwater
flux $E$ to the Indo-Pacific through the atmosphere; this is received by the
Indian and Pacific reservoirs in the proportion $\zeta$. Hence
\begin{equation}
    f_{n} \equiv h,
    \quad
    f_{t} \equiv -h-E,
    \quad
    f_{i} \equiv \zeta E,
    \quad
    f_{c} \equiv (1-\zeta)E,
    \quad
    f_{d} \equiv 0,
    \label{eq:forcings}
\end{equation}
so that $\sum_{j} f_{j} = 0$, which is the statement that the global
hydrological cycle conserves freshwater. Forming the volume-weighted sum of
Eqs.~(\ref{eq:box_n})--(\ref{eq:box_d}), the coefficient multiplying each
salinity is
\begin{align}
    s_{t}&: \quad Q - Q = 0, \qquad\qquad\ \,
    s_{n}: \quad -Q + Q = 0,
    \nonumber\\
    s_{i}&: \quad \gamma Q - \left(\gamma Q-\phi\right) - \phi - R + R = 0,
    \qquad
    s_{c}: \quad \phi - \phi = 0,
    \label{eq:cancel}\\
    s_{d}&: \quad (1-\gamma)Q + \left(\gamma Q-\phi\right)
      + \phi + R - Q - R = 0,
    \nonumber
\end{align}
so that every advective, throughflow, and retroflective contribution
cancels identically and
\begin{equation}
    \frac{\mathrm{d}}{\mathrm{d}\tau}\,\mathcal{S} = -\sum_{j} f_{j} = 0,
    \qquad
    \mathcal{S} \equiv \sum_{j} v_{j}s_{j}.
    \label{eq:invariant}
\end{equation}
The volume-weighted total salt is therefore an exact invariant of the flow.
This has two uses. Its drift under time integration measures the integrator
rather than the model, since the invariant is nowhere imposed. And because
the Jacobian of Eqs.~(\ref{eq:box_n})--(\ref{eq:box_d}) annihilates the
direction $\partial/\partial\mathcal{S}$ and is consequently singular,
eliminating one reservoir through
\begin{equation}
    s_{d} = \frac{1}{v_{d}}\left(\mathcal{S} - v_{n}s_{n} - v_{t}s_{t}
      - v_{i}s_{i} - v_{c}s_{c}\right)
    \label{eq:eliminate}
\end{equation}
removes the corresponding zero eigenvalue and renders the reduced
four-dimensional equilibrium problem nonsingular. Writing
$\mathbf{u} = (s_{n},s_{t},s_{i},s_{c})^{\top}$ and denoting the reduced
right-hand side by $\mathbf{G}(\mathbf{u},h)$, the chain rule applied to
Eq.~(\ref{eq:eliminate}) contributes
$\partial s_{d}/\partial u_{k} = -v_{k}/v_{d}$ to every row of the reduced
Jacobian $\partial\mathbf{G}/\partial\mathbf{u}$, which is assembled
analytically. Because the hosing enters only through $f_{n}$ and $f_{t}$,
the parameter derivative is the constant vector
$\partial\mathbf{G}/\partial h
= \left(-v_{n}^{-1},\,v_{t}^{-1},\,0,\,0\right)^{\top}$.

A consequence of Eq.~(\ref{eq:cancel}) that is central to what follows
concerns the gateway reservoirs alone. Adding the steady-state forms of
Eqs.~(\ref{eq:box_i}), (\ref{eq:box_c}), and (\ref{eq:box_d}) weighted by
their volumes, every exchange internal to the set $\{i,c,d\}$ appears twice
with opposite sign and cancels, exactly as in the full sum. The subsystem
communicates with the Atlantic only through the overturning, importing
$Q s_{n}$ from the North Atlantic and exporting
$\gamma Q s_{i} + (1-\gamma)Q s_{d}$ to the upper limb, and it receives the
net forcing $f_{i}+f_{c}+f_{d} = E$. The steady balance is therefore
\begin{equation}
    Q\left(s_{\mathrm{route}} - s_{n}\right) = E,
    \qquad
    s_{\mathrm{route}} \equiv \gamma s_{i} + (1-\gamma)s_{d},
    \label{eq:identity}
\end{equation}
in which the warm-route fraction $\gamma$, the throughflow $\phi$, the
retroflection $R$, and the apportionment $\zeta$ have all disappeared. Their
absence is not an approximation but a consequence of conservation: the
internal structure of the gateway determines how salt is distributed among
its reservoirs, whereas the export to the Atlantic is fixed by the net
forcing the subsystem receives. Substituting Eq.~(\ref{eq:identity}) into
Eq.~(\ref{eq:box_t}) gives
$v_{t}\dot{s}_{t} = -Q\left(s_{t}-s_{n}\right) + E - f_{t}$, so the gateway
contributes an additive constant to the upper-limb budget and cannot alter
the coefficient of the nonlinear term through which the salt-advection
feedback acts.

The formulation above assumes that a salinity anomaly leaving the deep
reservoir reaches the upper limb instantaneously. Advective transit through
the gateway in fact occupies a finite time, and because anomalies follow
many paths that time is distributed rather than concentrated at a single
lag. We therefore replace the instantaneous gateway signal by a memory
average against a normalized kernel,
\begin{equation}
    z(\tau) \equiv \int_{0}^{\infty} g_{n}(\sigma)\,x(\tau-\sigma)\,
      \mathrm{d}\sigma,
    \qquad
    g_{n}(\sigma) \equiv \frac{\left(n/\bar{\tau}\right)^{n}
      \sigma^{\,n-1}\mathrm{e}^{-n\sigma/\bar{\tau}}}{\Gamma(n)},
    \label{eq:kernel}
\end{equation}
where $\Gamma$ is the gamma function, $\bar{\tau}$ the mean transit time,
and the integer shape $n$ controls the dispersion: the variance of
$g_{n}$ is $\bar{\tau}^{2}/n$, so $n = 1$ gives a monotonically decaying
exponential memory and $n \to \infty$ recovers a discrete lag,
$g_{n} \to \delta(\sigma-\bar{\tau})$. Because the kernel is normalized,
$z = x$ at any steady state and the equilibria are unchanged by this
substitution.

For integer $n$ the convolution admits an exact finite-dimensional
representation. Define the partial averages
$z_{k}(\tau) \equiv \int_{0}^{\infty}g_{k}(\sigma)x(\tau-\sigma)\mathrm{d}\sigma$
for $k = 1,\ldots,n$, each formed with the common rate $n/\bar{\tau}$, so
that $z_{n} = z$. The Erlang kernels obey the recursion
$\mathrm{d}g_{1}/\mathrm{d}\sigma = -(n/\bar{\tau})g_{1}$ with
$g_{1}(0) = n/\bar{\tau}$, and
$\mathrm{d}g_{k}/\mathrm{d}\sigma
= (n/\bar{\tau})\left(g_{k-1}-g_{k}\right)$ with $g_{k}(0) = 0$ for
$k \geq 2$. Differentiating $z_{k}$ under the integral sign and applying
these relations yields the linear chain trick \cite{MacDonald1978,
Hurtado2019}:
\begin{equation}
    \dot{z}_{1} = \frac{n}{\bar{\tau}}\left(x-z_{1}\right),
    \qquad
    \dot{z}_{k} = \frac{n}{\bar{\tau}}\left(z_{k-1}-z_{k}\right),
    \quad k = 2,\ldots,n.
    \label{eq:chain_z}
\end{equation}
Letting the delayed gateway signal modulate the salt-advection gain with
sensitivity $\kappa$, the state
$\mathbf{y} = \left(x,z_{1},\ldots,z_{n}\right)^{\top}$ evolves according to
\begin{equation}
    \dot{x} = p - x\left[1 + \mu^{2}\left(1+\kappa z_{n}\right)
      (1-x)^{2}\right]
    \label{eq:chain_x}
\end{equation}
together with Eq.~(\ref{eq:chain_z}). The system is an ODE in
$\mathbb{R}^{n+1}$ that is exactly equivalent to the
integro-differential formulation, not an approximation to it.

Linearizing about an equilibrium $\left(x_{\ast},z_{\ast}\right)$ with
$z_{\ast} = x_{\ast}$, and defining the instantaneous and delayed feedback
coefficients
\begin{equation}
    a \equiv \left.\frac{\partial\dot{x}}{\partial x}\right|_{\ast}
      = -1 + \mu^{2}\left(1+\kappa z_{\ast}\right)
        \left(1-x_{\ast}\right)\left(3x_{\ast}-1\right),
    \qquad
    b \equiv \left.\frac{\partial\dot{x}}{\partial z_{n}}\right|_{\ast}
      = -x_{\ast}\mu^{2}\kappa\left(1-x_{\ast}\right)^{2},
    \label{eq:ab}
\end{equation}
the chain acts as a cascade of $n$ identical first-order low-pass filters
with transfer function
$\hat{g}_{n}(\lambda) = \left(1+\lambda\bar{\tau}/n\right)^{-n}$, so the
eigenvalue $\lambda$ satisfies
$\lambda - a - b\,\hat{g}_{n}(\lambda) = 0$. Clearing the denominator gives
the polynomial characteristic equation
\begin{equation}
    \left(\lambda-a\right)
      \left(1+\frac{\lambda\bar{\tau}}{n}\right)^{n} - b = 0
    \label{eq:charpoly}
\end{equation}
of degree $n+1$. As $n \to \infty$ the factor
$\left(1+\lambda\bar{\tau}/n\right)^{-n}$ tends to
$\mathrm{e}^{-\lambda\bar{\tau}}$ and Eq.~(\ref{eq:charpoly}) approaches the
transcendental discrete-delay equation
$\lambda - a - b\,\mathrm{e}^{-\lambda\bar{\tau}} = 0$, the approach
occurring at a rate linear in $1/n$ \cite{Ando2020}. The limiting problem
is infinite dimensional, its spectrum being fixed by a transcendental
characteristic function \cite{Diekmann1995}, whereas every member of the
family at finite $n$ is finite dimensional. Setting
$\lambda = \mathrm{i}\omega$ in Eq.~(\ref{eq:charpoly}) and writing
$u = \omega\bar{\tau}/n$, the modulus and argument of the two factors
separate. With $a < 0$ at a stable equilibrium, the factor $\mathrm{i}\omega
- a$ has modulus $\sqrt{a^{2}+\omega^{2}}$ and argument
$\arctan\left(\omega/|a|\right)$, while $\left(1+\mathrm{i}u\right)^{n}$ has
modulus $\left(1+u^{2}\right)^{n/2}$ and argument $n\arctan(u)$. Requiring
their product to equal the real number $b$ gives
\begin{equation}
    \arctan\!\left(\frac{\omega}{|a|}\right) + n\arctan(u) = \pi,
    \qquad
    b = -\sqrt{a^{2}+\omega^{2}}\left(1+u^{2}\right)^{n/2},
    \label{eq:phase}
\end{equation}
the first being a phase balance and the second fixing the delayed feedback
strength at which a complex-conjugate pair crosses the imaginary axis.

Each arctangent in Eq.~(\ref{eq:phase}) is bounded by $\pi/2$, so their sum
cannot reach $\pi$ when $n = 1$, and no delay-induced Hopf bifurcation
exists at any positive mean transit time under an exponential memory. For
$n \geq 2$ the least delayed feedback capable of destabilizing the
equilibrium follows by minimizing $|b|$ over $\bar{\tau}$. Setting
$\bar{\tau} = n/|a|$ makes $u = \omega\bar{\tau}/n = \omega/|a|$, so the two
arctangents in the phase balance become equal and
$(n+1)\arctan\left(\omega/|a|\right) = \pi$. The critical frequency and
threshold then follow in closed form,
\begin{equation}
    \omega^{\star} = |a|\tan\!\left(\frac{\pi}{n+1}\right),
    \qquad
    \frac{|b_{c}|}{|a|}
      = \sec^{\,n+1}\!\left(\frac{\pi}{n+1}\right),
    \qquad n \geq 2,
    \label{eq:threshold}
\end{equation}
the $n = 1$ case being recovered as the divergence of
$\sec(\pi/2)$. Writing $\vartheta = \pi/(n+1)$ and expanding
$\sec\vartheta = 1 + \vartheta^{2}/2 + \mathcal{O}(\vartheta^{4})$ gives the
large-shape behavior
\begin{equation}
    \frac{|b_{c}|}{|a|}
      = \exp\!\left[\frac{\pi^{2}}{2(n+1)}\right]
        \left(1+\mathcal{O}\!\left(n^{-2}\right)\right)
      = 1 + \frac{\pi^{2}}{2n} + \mathcal{O}\!\left(n^{-2}\right),
    \qquad \frac{\pi^{2}}{2} = 4.93,
    \label{eq:asymptotic}
\end{equation}
so the threshold approaches the discrete-delay value of unity from above at
a rate linear in $1/n$, consistent with the convergence rate of the
chain-trick representation \cite{Ando2020}. That $\bar{\tau} = n/|a|$ is the
minimizer rather than a stationary point of another kind is confirmed
numerically below.

Finally, fluctuations in the hydrological cycle perturb the forcing
directly, which makes the noise additive, so that the It\^{o} and
Stratonovich readings of the resulting equation coincide \cite{Cessi1994}:
\begin{equation}
    \mathrm{d}x = \left[p - x\left(1+m(1-x)^{2}\right)\right]\mathrm{d}\tau
      + \varsigma\,\mathrm{d}W,
    \label{eq:sde}
\end{equation}
with $\varsigma$ the noise amplitude and $W$ a standard Wiener process. The
drift is the negative gradient of the quartic potential
\begin{equation}
    U(x) \equiv -px + \frac{x^{2}}{2}
      + m\left(\frac{x^{2}}{2} - \frac{2x^{3}}{3}
      + \frac{x^{4}}{4}\right),
    \label{eq:potential}
\end{equation}
obtained by integrating $-f$, and the associated stationary density is
$p_{s}(x) \propto \exp\!\left[-2U(x)/\varsigma^{2}\right]$.

\subsection{Numerical implementation}
\label{subsec:numerics}

All computations were carried out in Python 3.12 with NumPy 2.4
\cite{Harris2020}, SciPy 1.17 \cite{Virtanen2020}, and Matplotlib 3.10
\cite{Hunter2007}. Figure palettes follow a colorblind-safe
qualitative scheme \cite{Wong2011}. No external data were used; every
quantity analyzed was generated by the procedures described below.

Equilibrium branches were obtained by pseudo-arclength continuation rather
than by advancing the forcing and integrating forward. A forward sweep
conflates the bifurcation structure with the rate at which the parameter is
advanced, and cannot follow the unstable branch at all, whereas continuation
treats the state and the parameter symmetrically \cite{Allgower2003,
Kuznetsov2004}; the predictor-corrector scheme adopted here is the one
underlying standard bifurcation software \cite{Dhooge2003}. For
the scalar problem, let $\mathbf{w} = (x,p)^{\top}$ and let
$\mathbf{w}_{0}$ and $\mathbf{t} = (t_{x},t_{p})^{\top}$ denote the previous
accepted point and the unit tangent there. A new point is sought as the root
of the extended system
\begin{equation}
    \mathbf{\Phi}(\mathbf{w}) \equiv
    \begin{pmatrix}
      f(x;p,m) \\[2pt]
      t_{x}\left(x-x_{0}\right) + t_{p}\left(p-p_{0}\right) - \Delta s
    \end{pmatrix}
    = \mathbf{0},
    \label{eq:extended}
\end{equation}
where $\Delta s$ is the arclength step. The Newton correction solves
$\mathbf{J}_{\mathrm{ext}}\,\delta\mathbf{w} = -\mathbf{\Phi}$ with
\begin{equation}
    \mathbf{J}_{\mathrm{ext}} \equiv
    \begin{pmatrix} f_{x} & f_{p} \\ t_{x} & t_{p} \end{pmatrix},
    \qquad
    \det\mathbf{J}_{\mathrm{ext}} = f_{x}t_{p} - f_{p}t_{x},
    \label{eq:bordered}
\end{equation}
which is inverted in closed form for the two-dimensional case. The
determinant remains bounded away from zero at a fold, where $f_{x} = 0$ but
$t_{x} = \pm 1$, which is precisely why the extended system succeeds where a
naive Newton iteration on $f$ alone fails. The tangent is the normalized
null vector of $\left(f_{x}\ \ f_{p}\right)$, namely
$\mathbf{t} \propto \left(-f_{p},\,f_{x}\right)^{\top}$, oriented by
requiring a positive inner product with the previous tangent so that the
branch is traversed monotonically. Branches were seeded at
$(x,p) = (0.02,0.02)$, Newton-projected onto the manifold, and continued
with $\Delta s = 1.50\times10^{-3}$ and corrector tolerance $10^{-12}$ for
$m \in \{2.40,\,3.00,\,4.00,\,6.00\}$.

For the four-dimensional reduced box system the same construction applies
with $\mathbf{w} = (\mathbf{u},h)$. The bordered matrix
\begin{equation}
    \mathbf{M} \equiv
    \begin{pmatrix}
      \partial\mathbf{G}/\partial\mathbf{u} &
      \partial\mathbf{G}/\partial h \\[2pt]
      \mathbf{t}_{\mathbf{u}}^{\top} & t_{h}
    \end{pmatrix}
    \in \mathbb{R}^{5\times5}
    \label{eq:bordered_vec}
\end{equation}
is factorized once per corrector iteration and reused to obtain the updated
tangent as the solution of
$\mathbf{M}\,\mathbf{t} = \mathbf{e}_{5}$, subsequently normalized. Branches
were continued in $h$ from $h = 0.02$ with
$\Delta s = 2.00\times10^{-3}$ and tolerance $10^{-11}$ for
$\gamma \in \{0.150,\,0.400,\,0.650,\,0.900\}$, with the leading real part
of the spectrum of $\partial\mathbf{G}/\partial\mathbf{u}$ recorded at every
point.

Folds were localized after continuation rather than during it. A sign change
of $f_{x}$ between consecutive branch points brackets a fold, which is then
refined by Newton iteration on the pair $f = f_{x} = 0$. Because $f_{x}$ is
independent of $p$, the Jacobian of this pair is block triangular and the
iteration reduces to the scalar update
\begin{equation}
    x \leftarrow x - \frac{f_{x}}{f_{xx}},
    \qquad
    p \leftarrow x\left[1+m(1-x)^{2}\right],
    \label{eq:foldnewton}
\end{equation}
iterated until $\max\left(|f|,|f_{x}|\right) < 10^{-14}$. At each converged
fold the quadratic normal-form coefficient $f_{xx}/2$ was evaluated; a
nonzero value certifies a genuine saddle-node rather than a transcritical
degeneracy, and localization through a minimally extended system of this
kind is the standard construction \cite{Kuznetsov2004, Dhooge2003}. Nine gains spanning
$m \in [3.20,\,8.80]$ were treated in this way and compared with
Eq.~(\ref{eq:folds}). For the box system, where no closed form is available,
folds were bracketed by sign changes of the leading real part and refined by
bisection in $h$ with the state reconverged by Newton at fixed parameter.

Analytic Jacobians were verified against complex-step derivatives
\cite{Squire1998}. Expanding an analytic $f$ about a real argument along the
imaginary direction,
\begin{equation}
    f(x+\mathrm{i}\eta) = f(x) + \mathrm{i}\eta f_{x}
      - \frac{\eta^{2}}{2}f_{xx}
      - \mathrm{i}\frac{\eta^{3}}{6}f_{xxx} + \mathcal{O}(\eta^{4}),
    \qquad
    \frac{\operatorname{Im}\left[f(x+\mathrm{i}\eta)\right]}{\eta}
      = f_{x} - \frac{\eta^{2}}{6}f_{xxx} + \mathcal{O}(\eta^{4}),
    \label{eq:cstep}
\end{equation}
so the estimate is second-order accurate and, crucially, involves no
difference of nearly equal quantities. Its error therefore decreases
monotonically with $\eta$ down to the point where the imaginary part
underflows, in contrast to a central difference
$\left[f(x+\eta)-f(x-\eta)\right]/2\eta$, whose total error is bounded below
by the competition between an $\mathcal{O}(\eta^{2})$ truncation term and an
$\mathcal{O}(\epsilon_{M}/\eta)$ cancellation term. The comparison was
performed at $x = 0.420$, $p = 0.950$, $m = 4.00$ over sixty logarithmically
spaced values of $\eta$ from $10^{-1}$ to $10^{-18}$, with $\eta = 10^{-20}$
used in production.

Spectra of Eq.~(\ref{eq:charpoly}) were obtained by expanding the binomial
explicitly,
\begin{equation}
    \left(\lambda-a\right)\sum_{k=0}^{n}\binom{n}{k}
      \left(\frac{\bar{\tau}}{n}\right)^{k}\lambda^{k} - b = 0,
    \label{eq:expanded}
\end{equation}
and computing the roots of the resulting degree-$(n+1)$ polynomial as the
eigenvalues of its companion matrix through \texttt{numpy.roots}. This
converts what would otherwise be an infinite-dimensional stability problem
\cite{Diekmann1995} into a standard eigenvalue computation, and avoids the
root finding on a transcendental characteristic function that delayed
bifurcation software must carry out \cite{Engelborghs2002}. It is in this
respect an alternative to pseudospectral discretization of the delayed
generator \cite{Diekmann2020}, exact for the gamma kernel rather than
convergent in the discretization order.
Spectra were evaluated for $n \in \{2,\,8,\,16,\,32\}$.

The Hopf threshold was obtained from Eq.~(\ref{eq:phase}). For fixed $n$ and
$\bar{\tau}$ the phase balance is a monotone equation in $\omega$ whose root
was located by Brent's method \cite{Brent1973} through
\texttt{scipy.optimize.brentq} after bracketing by geometric expansion; the
corresponding $|b|$ then follows from the magnitude relation. Minimizing
$|b|$ over $\bar{\tau}$ at fixed $n$ gives the least delayed-feedback
strength for which memory can destabilize the equilibrium at any mean
transit time. The minimization was performed over a logarithmic grid of
$4000$ values of $\bar{\tau}$ spanning $10^{-3}$ to $10^{4}$, at $a = -1$
and $n \in \{1,2,\ldots,128\}$. Gamma kernels and their transfer moduli
$\left|\hat{g}_{n}(\omega)\right|
= \left(1+\left(\omega\bar{\tau}/n\right)^{2}\right)^{-n/2}$ were tabulated
for $n \in \{1,2,4,16,64\}$.

Stiff initial-value problems were integrated with the fifth-order implicit
Radau~IIA method \cite{Hairer1996} through
\texttt{scipy.integrate.solve\_ivp}, at relative tolerance $10^{-11}$ and
absolute tolerance $10^{-13}$ unless stated otherwise. The chain system was
integrated for $n \in \{2,16\}$ at $\mu^{2} = 2.40$ and $\kappa = 1.00$,
with $\bar{\tau} = 6/|a|$ and a common forcing, from an initial perturbation
of $0.02$ about the equilibrium over $\tau \in [0,260]$. Equilibria of the
chain and box systems were located with \texttt{scipy.optimize.brentq} where
a bracket was available and with the hybrid Powell method through
\texttt{scipy.optimize.fsolve} otherwise. Drift of the invariant
$\mathcal{S}$ was recorded over nine tolerances from $10^{-4}$ to
$10^{-12}$ for $\tau \in [0,60]$ from the state
$\mathbf{s}(0) = (0.05,\,0.30,\,0.28,\,0.06,\,0.02)^{\top}$, with baseline
box parameters $v_{n} = v_{t} = 1.00$, $v_{i} = 2.00$, $v_{c} = 4.00$,
$v_{d} = 8.00$, $\mu^{2} = 5.00$, $\gamma = 0.550$, $\phi = 0.600$,
$R = 0.800$, $E = 0.220$, and $\zeta = 0.600$. The gateway salt export
$Q\left(s_{\mathrm{route}}-s_{n}\right)$, with
$s_{\mathrm{route}} = \gamma s_{i} + (1-\gamma)s_{d}$, was evaluated at
$h = 0.250$ over $\gamma \in [0.05,\,0.98]$ and
$R \in \{0.200,\,0.800,\,2.00\}$.

Trajectories of Eq.~(\ref{eq:sde}) were generated with the Euler--Maruyama
scheme and with a Stratonovich Heun scheme,
\begin{align}
    X_{k+1} &= X_{k} + f\left(X_{k}\right)\Delta\tau
      + \varsigma\,\Delta W_{k},
    \label{eq:em}\\
    \tilde{X} &= X_{k} + f\left(X_{k}\right)\Delta\tau
      + \varsigma\,\Delta W_{k},
    \qquad
    X_{k+1} = X_{k} + \frac{\Delta\tau}{2}
      \left[f\left(X_{k}\right) + f\left(\tilde{X}\right)\right]
      + \varsigma\,\Delta W_{k},
    \label{eq:heun}
\end{align}
with $\Delta W_{k} \sim \mathcal{N}\left(0,\Delta\tau\right)$. Both attain
strong order one for additive noise \cite{Kloeden1992, Higham2001}, since
the Milstein correction vanishes when the diffusion coefficient is
constant.
Strong convergence was measured against a common reference path computed
with $2^{16}$ Euler--Maruyama steps on $\tau \in [0,1]$. Coarse increments
were formed as exact sums of the fine increments they span,
\begin{equation}
    \Delta W_{j}^{(N)} \equiv \sum_{l=jM+1}^{(j+1)M}\Delta W_{l}^{(2^{16})},
    \qquad
    M \equiv 2^{16}/N,
    \label{eq:brownian}
\end{equation}
so that all schemes at all step sizes are driven by the same realization of
the Wiener process and the comparison isolates discretization error.
Ensembles of $400$ realizations were run at
$N \in \{2^{4},\ldots,2^{9}\}$ with $\varsigma = 0.150$, $m = 4.00$, and
$p = 0.950$.

Diagnostics were defined directly on this output. Stability of an
equilibrium was assessed from $\max_{k}\operatorname{Re}\lambda_{k}$,
obtained from the analytic Jacobian for the equilibrium problems and from
Eq.~(\ref{eq:expanded}) for the delayed problem; folds were identified as
sign changes of $f_{x}$, and Hopf bifurcations as imaginary-axis crossings
of a complex-conjugate pair. Accuracy was reported as the absolute error
against the closed-form values of Eqs.~(\ref{eq:folds}) and (\ref{eq:cusp}),
benchmarked against the unit roundoff $\epsilon_{M}\left|f_{x}\right|$ with
$\epsilon_{M} = 2.22\times10^{-16}$, and continuation residuals as
$\max|f|$ over all computed branch points. Drift of the salt invariant was
reported in relative form,
\begin{equation}
    \mathcal{D} \equiv \max_{\tau\in[0,60]}
      \frac{\left|\mathcal{S}(\tau)-\mathcal{S}(0)\right|}
           {\left|\mathcal{S}(0)\right|},
    \label{eq:drift}
\end{equation}
and departure from the gateway identity as
$\left|Q\left(s_{\mathrm{route}}-s_{n}\right)+E\right|$, reported as the
maximum over the sweep in $\gamma$ and $R$. Observed orders of strong
convergence were estimated as the ordinary least-squares slope
\begin{equation}
    \hat{\gamma}_{s} \equiv \frac{\operatorname{Cov}
      \left(\ln\Delta\tau,\,\ln\mathcal{E}\right)}
      {\operatorname{Var}\left(\ln\Delta\tau\right)},
    \qquad
    \mathcal{E}\left(\Delta\tau\right) \equiv \frac{1}{400}\sum_{r=1}^{400}
      \left|X_{T}^{\Delta\tau,\,r} - X_{T}^{\mathrm{ref},\,r}\right|,
    \label{eq:order}
\end{equation}
where $X_{T}$ denotes the terminal state of realization $r$. All numerical
values below are reported to three significant figures unless comparison
with machine precision requires otherwise.
\section{Results}
\label{sec:results}

\begin{figure}[H]
    \centering
    \includegraphics[width=\linewidth]{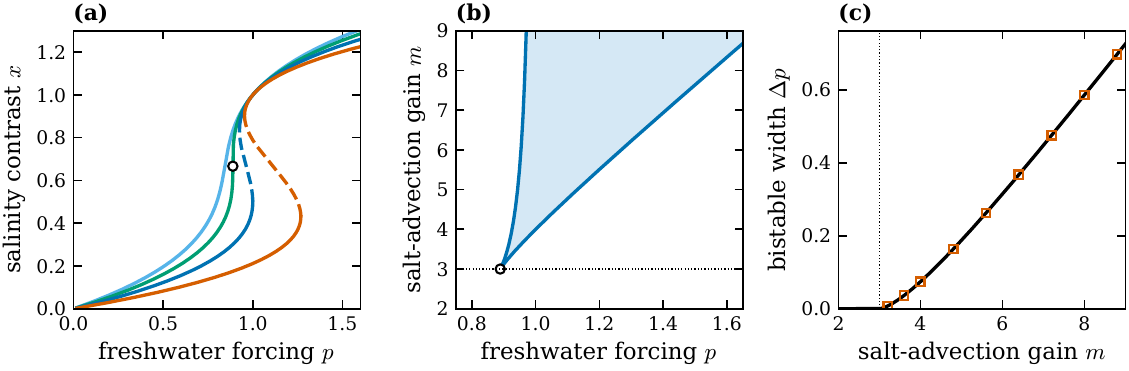}
    \caption{Equilibrium structure of the scalar model. (a) Branches
    continued at $m = 2.40$, $3.00$, $4.00$, and $6.00$; solid segments are
    stable and dashed segments unstable, and the open circle marks the exact
    cusp at $(p, x) = (8/9, 2/3)$. (b) Fold locus in the forcing-gain plane,
    closing at the cusp $m = 3$; the shaded region is the bistable set.
    (c) Bistable width against gain, comparing the closed form of
    Eq.~(\ref{eq:folds}) with folds located by the bordered Newton solve at
    nine gains.}
    \label{fig:hysteresis}
\end{figure}

The equilibrium structure of the scalar model is shown in
Fig.~\ref{fig:hysteresis}. Continuation of Eq.~(\ref{eq:scalar}) yields a
monotone equilibrium manifold for $m = 2.40$ and folded manifolds for
$m = 4.00$ and $m = 6.00$; at $m = 3.00$ the manifold is monotone with a
vertical tangent at the cusp. Continuation produced $1685$ branch points at
$m = 4.00$ with a maximum equilibrium residual $\max|f| =
9.99\times10^{-13}$ and at most three corrector iterations per point. The
cusp located by the bordered Newton solve coincides with
Eq.~(\ref{eq:cusp}) at $(m_{c}, x_{c}, p_{c}) = (3.00, 0.667, 0.889)$.

Exact fold positions from Eq.~(\ref{eq:folds}) and the corresponding
bistable widths are listed in Table~\ref{tab:folds}. The width increases
monotonically from $\Delta p = 7.41\times10^{-3}$ at $m = 3.20$ to
$\Delta p = 0.698$ at $m = 8.80$. Folds located numerically agree with the
closed form to a maximum absolute error of $2.78\times10^{-15}$ in the state
and $2.22\times10^{-16}$ in the forcing across the nine gains tested, with
residuals $\max|f| = 0$ and $\max|f_{x}| \leq 8.33\times10^{-15}$ and at
most four corrector iterations. At $m = 4.00$ the quadratic normal-form
coefficients are $f_{xx}/2 = 2.00$ at the lower fold and $-2.00$ at the
upper fold.

\begin{table}[H]
\centering
\caption{Exact fold states and bistable widths of the scalar model from
Eqs.~(\ref{eq:folds}) and (\ref{eq:scalar}).}
\label{tab:folds}
\small
\begin{tabular}{cccc}
\toprule
Gain $m$ & $x_{-}$ & $x_{+}$ & Width $\Delta p$ \\
\midrule
3.20 & 0.583 & 0.750 & $7.41\times10^{-3}$ \\
3.60 & 0.531 & 0.803 & $3.63\times10^{-2}$ \\
4.00 & 0.500 & 0.833 & $7.41\times10^{-2}$ \\
4.80 & 0.463 & 0.871 & 0.163 \\
5.60 & 0.440 & 0.894 & 0.262 \\
6.40 & 0.424 & 0.910 & 0.367 \\
7.20 & 0.412 & 0.921 & 0.475 \\
8.00 & 0.403 & 0.930 & 0.586 \\
8.80 & 0.396 & 0.937 & 0.698 \\
\bottomrule
\end{tabular}
\end{table}

\begin{figure}[H]
    \centering
    \includegraphics[width=\linewidth]{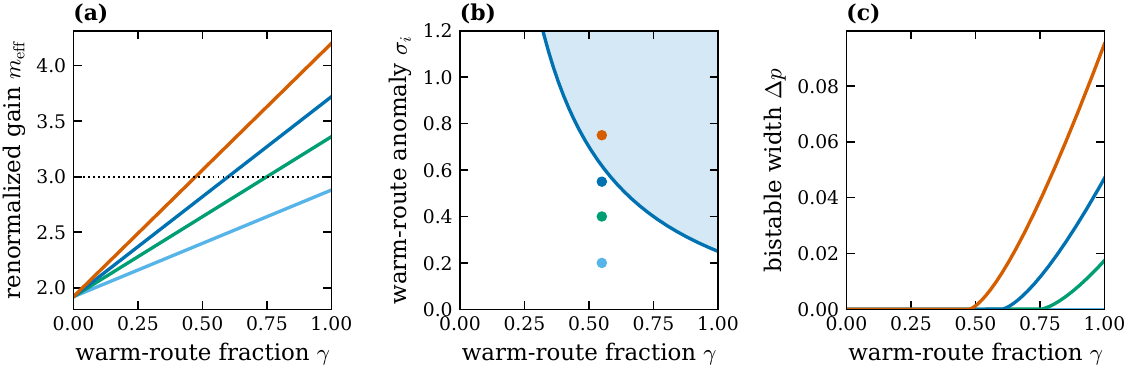}
    \caption{Consequences of the hypothesized renormalization of the
    salt-advection gain. (a) Renormalized gain against warm-route fraction
    for four warm-route source anomalies, with the cusp gain $m = 3$ marked.
    (b) Marginal-stability boundary in the plane of warm-route fraction and
    warm-route source anomaly; the shaded region would be bistable. Circles
    mark the four parameter sets of panel (a) at $\gamma = 0.550$.
    (c) Bistable width that would follow, closing continuously at the
    boundary.}
    \label{fig:gateway}
\end{figure}

Under the hypothesis that interbasin exchange renormalizes the gain
(Fig.~\ref{fig:gateway}), the renormalized gain increases linearly with the
warm-route fraction at each warm-route source anomaly tested. At the
baseline parameters $m_{0} = 2.40$, $\kappa = 1.00$, $\sigma_{i} = 0.550$,
$\sigma_{s} = -0.200$, and $\gamma = 0.550$, the route-averaged anomaly is
$0.213$ and the renormalized gain is $m_{\mathrm{eff}} = 2.91$, which lies
below the cusp value $3$, so that the corresponding bistable width is zero.
The marginal-stability boundary in the $(\gamma, \sigma_{i})$ plane is a
rectangular hyperbola, and the predicted bistable width rises from zero at
the boundary with the square-root scaling characteristic of a cusp.

\begin{figure}[H]
    \centering
    \includegraphics[width=\linewidth]{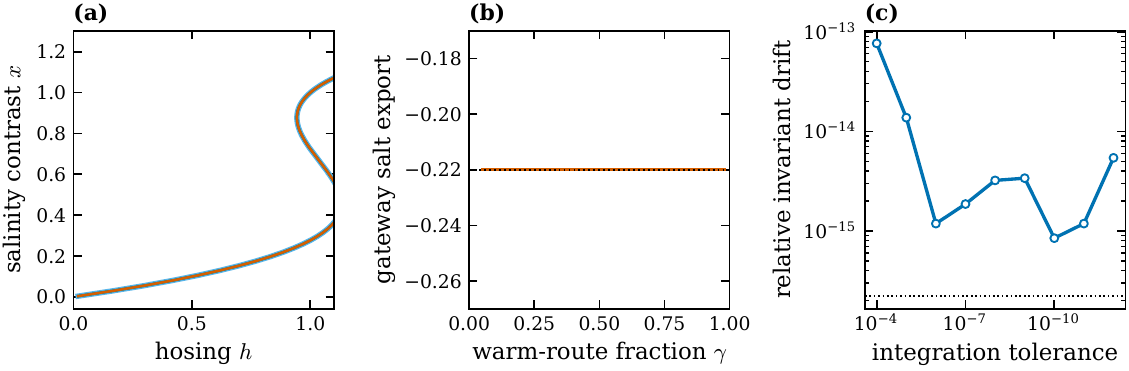}
    \caption{Closed salt budget of the five-box system. (a) Equilibrium
    branches continued in the hosing $h$ at four warm-route fractions
    $\gamma \in \{0.150, 0.400, 0.650, 0.900\}$; the curves coincide.
    (b) Gateway salt export $Q\,(s_{\mathrm{route}}-s_{n})$ against
    warm-route fraction at three retroflection strengths, together with the
    value $-E$ predicted by Eq.~(\ref{eq:identity}). (c) Relative drift of
    the volume-weighted total salt under Radau~IIA integration against
    tolerance, with the unit roundoff marked.}
    \label{fig:budget}
\end{figure}

The closed salt budget of the five-box system is evaluated in
Fig.~\ref{fig:budget}. The identity of Eq.~(\ref{eq:identity}) was evaluated numerically at $h = 0.250$ and $E = 0.220$, the gateway salt
export equals $-0.220$ at every combination of
$\gamma \in \{0.150, 0.400, 0.650, 0.900\}$ and
$R \in \{0.200, 0.800, 2.00\}$, with departures ranging from
$8.48\times10^{-14}$ to $1.81\times10^{-11}$ and a maximum of
$1.81\times10^{-11}$ over the twelve combinations. Equilibrium branches of
the five-box system continued in $h$ coincide for all four warm-route
fractions. The virtual salt fluxes of Eq.~(\ref{eq:forcings}) sum to
$-1.39\times10^{-17}$ and the instantaneous drift of the invariant is
$5.55\times10^{-17}$ at a reference total salt of $\mathcal{S} = 1.31$.
Under Radau~IIA integration the relative drift of $\mathcal{S}$ falls from
$7.64\times10^{-14}$ at tolerance $10^{-4}$ to between $8.47\times10^{-16}$
and $5.42\times10^{-15}$ for tolerances of $10^{-6}$ and below.

\begin{figure}[H]
    \centering
    \includegraphics[width=\linewidth]{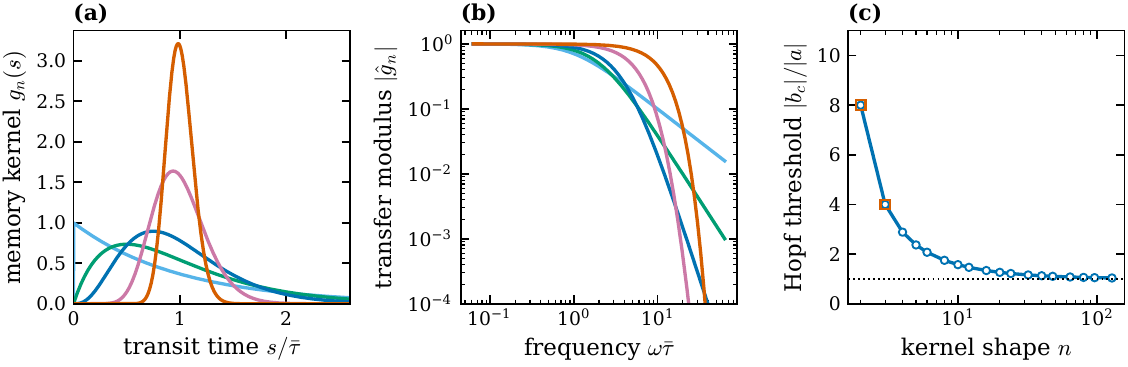}
    \caption{Memory dispersion and the delay-induced Hopf threshold.
    (a) Gamma kernels of common mean $\bar{\tau}$ and shapes
    $n \in \{1, 2, 4, 16, 64\}$. (b) Modulus of the kernel transfer
    function for the same shapes. (c) Minimum delayed-feedback strength
    admitting a Hopf bifurcation against kernel shape (circles), with the
    exact values $8|a|$ at $n = 2$ and $4|a|$ at $n = 3$ (squares) and the
    discrete-delay asymptote $|b| = |a|$ (dotted). No solution exists at
    $n = 1$.}
    \label{fig:memory}
\end{figure}

Properties of the gamma memory kernel are shown in
Fig.~\ref{fig:memory}. At fixed mean transit time the kernel narrows as the
shape increases, from a monotonically decaying exponential at $n = 1$ to a
distribution peaked near $\sigma = \bar{\tau}$ at $n = 64$, and the transfer
modulus $|\hat{g}_{n}(\omega)|$ steepens correspondingly at high frequency.
Numerically computed thresholds agree with Eq.~(\ref{eq:threshold}) to a
maximum relative difference of $5.63\times10^{-7}$ over
$n \in [2,128]$, and the numerically located optimum satisfies
$\bar{\tau}^{\star}|a|/n = 1.00$ to within $1.30\times10^{-3}$. Values are listed in Table~\ref{tab:thresholds}, and the
asymptotic form of Eq.~(\ref{eq:asymptotic}) is approached from above. The computed quantity $n\left(|b_{c}|/|a| - 1\right)$ decreases from
$14.0$ at $n = 2$ to $4.99$ at $n = 128$. The discrete-delay limit
$n \to \infty$ gives $|b_{c}|/|a| = 1.00$.

\begin{table}[H]
\centering
\caption{Delay-induced Hopf threshold against memory kernel shape at
$a = -1$. Numerical values are obtained from Eq.~(\ref{eq:phase}); exact
values are from Eq.~(\ref{eq:threshold}).}
\label{tab:thresholds}
\small
\begin{tabular}{ccccc}
\toprule
Shape $n$ & $|b_{c}|/|a|$ (numerical) & $|b_{c}|/|a|$ (exact) &
$\bar{\tau}^{\star}|a|$ & $n\left(|b_{c}|/|a|-1\right)$ \\
\midrule
1 & none & divergent & --- & --- \\
2 & 8.000 & 8.000 & 2.00 & 14.0 \\
3 & 4.000 & 4.000 & 2.99 & 9.00 \\
4 & 2.885 & 2.885 & 4.00 & 7.54 \\
6 & 2.075 & 2.075 & 5.99 & 6.45 \\
8 & 1.750 & 1.750 & 8.01 & 6.00 \\
12 & 1.467 & 1.467 & 12.0 & 5.61 \\
16 & 1.339 & 1.339 & 16.0 & 5.43 \\
24 & 1.219 & 1.219 & 24.0 & 5.25 \\
32 & 1.162 & 1.162 & 32.0 & 5.17 \\
48 & 1.106 & 1.106 & 47.9 & 5.09 \\
64 & 1.079 & 1.079 & 64.1 & 5.05 \\
96 & 1.052 & 1.052 & 95.9 & 5.01 \\
128 & 1.039 & 1.039 & 128 & 4.99 \\
$\infty$ & 1.000 & 1.000 & --- & 4.93 \\
\bottomrule
\end{tabular}
\end{table}

\begin{figure}[H]
    \centering
    \includegraphics[width=\linewidth]{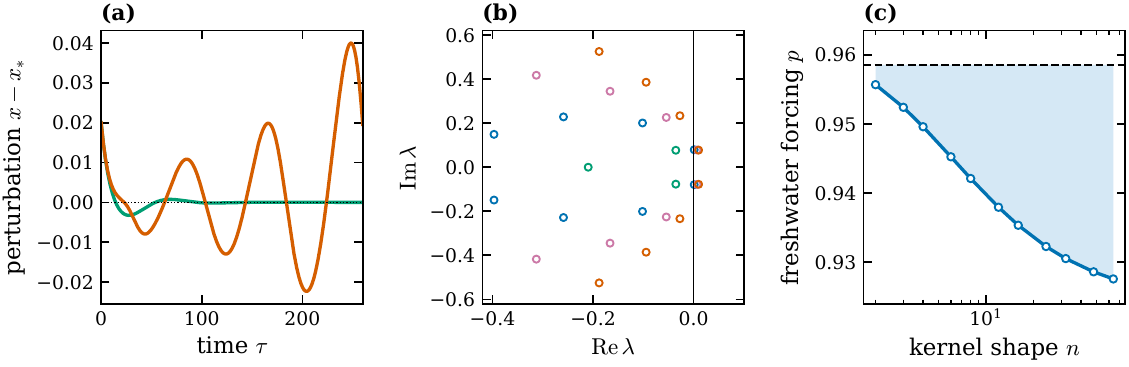}
    \caption{Delay-induced dynamics. (a) Trajectories of the linear-chain
    system at identical mean transit time and forcing for a broad kernel
    ($n = 2$) and a sharp kernel ($n = 16$). (b) Spectra of the
    characteristic polynomial Eq.~(\ref{eq:charpoly}) at the same operating
    point for $n \in \{2, 8, 16, 32\}$; the vertical line is the imaginary
    axis. (c) Forcing at the Hopf bifurcation against kernel shape
    (circles) together with the fold (dashed); the shaded interval is the
    forcing range over which a steady overturning persists.}
    \label{fig:dynamics}
\end{figure}

Dynamic consequences of the memory kernel are shown in
Fig.~\ref{fig:dynamics}. At identical mean transit time and forcing, a
perturbation of $0.02$ about the equilibrium relaxes monotonically for
$n = 2$ and grows into a sustained oscillation for $n = 16$. The spectra of
Eq.~(\ref{eq:charpoly}) at the same operating point show the leading
complex-conjugate pair migrating toward and across the imaginary axis as the
shape increases through $n \in \{2, 8, 16, 32\}$. On the lower branch of the
five-box system the instantaneous feedback $a$ vanishes at the fold, located
at $x_{\mathrm{fold}} = 0.525$ and $p_{\mathrm{fold}} = 0.959$, so that the
ratio $|b/a|$ increases without bound as the fold is approached and the
threshold of Eq.~(\ref{eq:threshold}) is met at a forcing strictly below the
saddle-node for every $n \geq 2$. The Hopf forcing decreases monotonically
from $p_{H} = 0.956$ at $n = 2$ to $p_{H} = 0.928$ at $n = 64$,
corresponding to intervals $p_{\mathrm{fold}} - p_{H}$ of
$2.88\times10^{-3}$ and $3.10\times10^{-2}$ respectively, or $0.300\%$ and
$3.23\%$ of the fold forcing, a ratio of $10.8$ between the two extremes.
Intermediate values are $1.72\%$ at $n = 8$, $2.42\%$ at $n = 16$, and
$2.92\%$ at $n = 32$.

\begin{figure}[H]
    \centering
    \includegraphics[width=\linewidth]{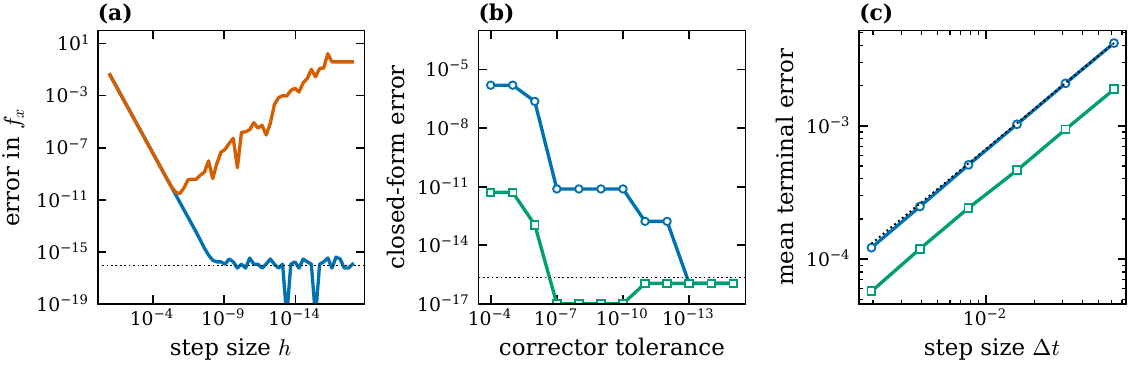}
    \caption{Numerical verification. (a) Error in $f_{x}$ against step size
    for the complex-step estimate and a central difference, with the unit
    roundoff marked. (b) Error of located folds against the closed form of
    Eq.~(\ref{eq:folds}) as the corrector tolerance is tightened.
    (c) Strong convergence of the stochastic integrators against a common
    fine reference path, with an order-one reference line.}
    \label{fig:verification}
\end{figure}

Verification diagnostics are collected in Fig.~\ref{fig:verification} and
Table~\ref{tab:verification}. At $x = 0.420$, $p = 0.950$, and $m = 4.00$
the analytic derivative is $f_{x} = -0.397$ and the complex-step estimate
agrees to $5.55\times10^{-17}$, against a unit roundoff on this magnitude of
$8.81\times10^{-17}$; the complex-step error is independent of step size
below $\eta \approx 10^{-8}$, whereas the central difference attains a
minimum error near $\eta \approx 10^{-6}$ and degrades on either side. Fold
localization errors against the closed form fall to $1.11\times10^{-16}$ at
corrector tolerances of $10^{-13}$ and below. The observed order of strong
convergence is $1.02$ for Euler--Maruyama and $1.00$ for the Heun scheme,
with terminal errors decreasing from $4.18\times10^{-3}$ and
$1.88\times10^{-3}$ at $\Delta\tau = 6.25\times10^{-2}$ to
$1.22\times10^{-4}$ and $5.73\times10^{-5}$ at
$\Delta\tau = 1.95\times10^{-3}$, an error ratio of $2.13$ at the finest
step.

\begin{table}[H]
\centering
\caption{Numerical verification summary. Errors are absolute unless
otherwise indicated.}
\label{tab:verification}
\small
\begin{tabular}{lc}
\toprule
Quantity & Value \\
\midrule
Analytic against complex-step $f_{x}$ & $5.55\times10^{-17}$ \\
Unit roundoff on this magnitude & $8.81\times10^{-17}$ \\
Maximum continuation residual $\max|f|$ & $9.99\times10^{-13}$ \\
Maximum corrector iterations & 3 \\
Lower fold state error & $5.55\times10^{-17}$ \\
Upper fold state error & $1.11\times10^{-16}$ \\
Lower fold forcing error & $1.11\times10^{-16}$ \\
Upper fold forcing error & $1.11\times10^{-16}$ \\
Normal-form coefficient, lower fold & 2.00 \\
Normal-form coefficient, upper fold & $-2.00$ \\
Hopf threshold against Eq.~(\ref{eq:threshold}), maximum relative
  & $5.63\times10^{-7}$ \\
Gateway identity, maximum departure & $1.81\times10^{-11}$ \\
Salt invariant, relative drift & $\leq 5.42\times10^{-15}$ \\
Euler--Maruyama observed strong order & 1.02 \\
Heun observed strong order & 1.00 \\
\bottomrule
\end{tabular}
\end{table}

\section{Discussion}
\label{sec:discussion}

The identity of Eq.~(\ref{eq:identity}) appears to place a structural
constraint on how interbasin gateways can act within conservative box
models. Its derivation requires only that the gateway subsystem exchange
salt with the Atlantic exclusively through the overturning and that the
total salt be conserved; under those conditions the steady-state export
salinity of the subsystem is pinned to its import salinity plus a term set
by the net forcing alone. The route partition, the throughflow strength, the
retroflection, and the distribution of the net Atlantic freshwater export
among the gateway reservoirs all cancel. The numerical departure of at most
$1.81\times10^{-11}$ across the parameter sweep is consistent with roundoff
accumulated through the nonlinear solve rather than with any residual
parameter dependence, and the coincidence of the equilibrium branches in
Fig.~\ref{fig:budget}a is consistent with the same conclusion. The intuition
that a saltier warm route must strengthen the salt-advection feedback thus
appears to fail for a reason that is essentially kinematic: whatever
additional salt the warm route concentrates is carried by a proportionally
reduced volume, and mass conservation enforces the compensation exactly.

It is important not to overstate the reach of this result. It does not
contradict the substantial body of work documenting Agulhas influence on
Atlantic salinity and stability \cite{Beal2011, Weijer2019, LeBars2013},
because in GCMs the Indo-Atlantic exchange is not confined to the
overturning. Mesoscale ring shedding, eddy salt fluxes at
the retroflection, and the wind-driven supergyre all transport salt along
pathways that violate the premise of Eq.~(\ref{eq:identity}). The
representation of the retroflection is itself strongly resolution dependent,
since a viscous boundary layer at coarse resolution gives way to an inertial
regime, in which the leakage is choked by turbulence, only at
eddy-permitting resolution \cite{LeBars2012, LeBars2013}; the leakage
further responds inversely to the strength of the Agulhas Current itself
\cite{vanSebille2009}, a dependence that no box model reproduces. Our result is
therefore better read as a statement about what a well-mixed, conservative
box model can and cannot represent: any gateway mechanism that does affect
equilibrium bistability must operate through a channel that such a model
omits by construction. To the extent that this is correct, it may help
explain why box-model and general-circulation-model accounts of interbasin
influence have proven difficult to reconcile, and it suggests that
attributing bistability shifts to the route partition within a box framework
requires an explicit non-overturning pathway rather than a reapportionment
of the existing one. A loose parallel is available on the observational
side. Over the multidecadal record the statistically detectable coupling
among these systems is confined to the ITF-Agulhas leg and is concentrated
at annual periods, with no direct signal reaching the AMOC at
26$^{\circ}$N \cite{Herho2026}. The correspondence with
Eq.~(\ref{eq:identity}) should not be pressed far, since a statistical null
over four decades and a structural cancellation in a conservative box model
are claims of different kinds, and the former is at least as readily
attributed to the limited meridional coherence of the overturning
\cite{Weijer2019}. It is nonetheless consistent with the picture developed
here, in which what the gateway transmits to the Atlantic is transient and
band-limited rather than a displacement of the equilibrium structure.

The identity also bears on the interpretation of freshwater-transport
diagnostics. The minimum of the overturning-induced freshwater transport at
the Atlantic southern boundary has been developed as a physics-based
early-warning indicator and applied to reanalysis products
\cite{vanWesten2024}. Equation~(\ref{eq:identity}) is consistent with the
emphasis such diagnostics place on the net flux rather than on its
composition, since within this model class the net gateway forcing is the
only quantity that survives into the equilibrium problem. At the same time,
the overturning freshwater transport on its own has been found to be a poor
predictor of the critical hosing rate in a coupled model \cite{Wood2019}, which is difficult to reconcile with a purely additive gateway unless
mechanisms outside the present framework are operating. These two findings
need not conflict, but their coexistence suggests that the relationship
between net freshwater transport and threshold proximity is unlikely to be
captured by a single scalar diagnostic.

Turning to the delayed problem, the closed-form threshold of
Eq.~(\ref{eq:threshold}) makes precise a phenomenon that has been reported
qualitatively in other settings. That distributed delays are more
stabilizing than discrete delays of the same mean is established in
mathematical biology, where gamma kernels are standard and the linear chain
trick provides an exact finite-dimensional representation
\cite{Hurtado2019}; recent analyses of ecological and epidemiological models
have found the weak (exponential) kernel to be the most stabilizing and the
discrete delay the least \cite{Ando2020}. The present contribution is
therefore not the existence of the effect but its exact form in this
context, together with the observation that conceptual AMOC models
incorporating advective lags have to our knowledge been formulated
exclusively with discrete delays. If Eq.~(\ref{eq:threshold}) is accepted,
the discrete delay is the least stable member of a one-parameter family, and
oscillatory behavior obtained under that idealization may be partly an
artifact of it. The magnitude of the discrepancy is not small: a delayed
feedback eight times the instantaneous feedback is required to destabilize
the equilibrium under an exponential-type kernel with $n = 2$, against a
factor of unity in the discrete limit. The asymptotic form
$1 + \pi^{2}/(2n)$ indicates that this sensitivity decays only as $1/n$,
which is consistent with the linear convergence rate of the chain-trick
approximation established by And\`{o} et al. \cite{Ando2020} and implies
that a physically realistic kernel would need to be quite sharply peaked
before discrete-delay results become quantitatively reliable.

The erosion of the steady window reported in Fig.~\ref{fig:dynamics}c
follows from the divergence of $|b/a|$ at the fold and is therefore generic
rather than parameter specific. Its practical significance depends on
whether the oscillation that sets in below the saddle-node is supercritical,
in which case small-amplitude variability would be expected, or subcritical,
in which case the active state could be lost outright. We have computed
linear thresholds only and have not determined the criticality, so no claim
about the outcome at onset can be made here. This caveat carries weight,
because in a calibrated five-box formulation \cite{Wood2019} the loss of
stability of the active state has been shown to occur through a subcritical
Hopf bifurcation rather than a saddle-node \cite{Alkhayuon2019}, with consequences for the geometry of the basin of attraction
near the bifurcation. Quasipotential and geometric early-warning diagnostics
have since been developed for that same box model \cite{Chapman2025}, and
the shape of the hysteresis loop traced under stochastic forcing has been
shown to depend on the noise itself \cite{Cini2025}; both would require
revisiting under a distributed kernel. Should the same criticality hold under a distributed
kernel, the intervals reported here would delimit forcing ranges in which
the active state is vulnerable to finite perturbations well before the
classical threshold is reached. Establishing this would require computing
the first Lyapunov coefficient of the chain system, which the
finite-dimensional representation makes tractable using standard normal-form
machinery \cite{Kuznetsov2004} and which we regard as the most informative
next step.

A further limitation concerns the provenance of the kernel. In the present
formulation $\bar{\tau}$ and $n$ are imposed rather than derived, whereas
they ought in principle to be determined by the residence-time distribution
of the gateway subsystem and hence by $\gamma$, $\phi$, and $R$. Until that
link is made, the results of Fig.~\ref{fig:dynamics} describe the
consequences of a kernel family rather than of the Indo-Pacific gateway
specifically. This gap is consequential given Eq.~(\ref{eq:identity}):
because the equilibrium channel is closed, the memory kernel is the only
route by which the gateway can influence the system within this model, so
deriving the kernel from the box geometry would connect the two halves of
the analysis. Available transit-time estimates suggest that such a
calculation would be well constrained, since throughflow water is understood
to reach the Agulhas region and subsequently the leakage on decadal
timescales \cite{LeBars2013, Gordon1986}, but we have not attempted the
calculation and refrain from assigning values. A related extension concerns
forcing applied at a finite rate. Bifurcation-induced, noise-induced, and
rate-induced tipping are distinct mechanisms \cite{Ashwin2012}, the last of
which has been argued to threaten the overturning under sufficiently rapid
ice melt \cite{Lohmann2021} and now admits a threshold characterization in
terms of connecting orbits \cite{Wieczorek2023}. How a distributed memory
modifies such thresholds is not addressed here, though the
finite-dimensional representation would make the computation accessible.

The broader interpretive context remains unsettled, and we are cautious
about extrapolating from a conceptual model to it. Statistical
early-warning analyses of observational proxies have been read as indicating
loss of stability \cite{Boers2021}, and a physics-based indicator applied to
reanalysis has been interpreted similarly \cite{vanWesten2024}, whereas
Southern Ocean wind-driven upwelling has been reported to sustain a weakened
circulation across 34 models under extreme forcing, implying that complete
collapse is unlikely within the present century \cite{Baker2025}. Much of this divergence concerns processes, in particular the
Southern Ocean upwelling balance and the possible emergence of a
compensating Pacific overturning cell, that a five-box Atlantic-centered
model does not represent. Our results speak to none of these questions
directly. What they do suggest, tentatively, is that the widely
reproduced hysteresis of the conceptual framework \cite{Stommel1961,
Cessi1994, Rahmstorf2005} is less sensitive to the composition of the
interbasin return flow than has sometimes been supposed, and more sensitive
than has been recognized to the temporal structure of the advective feedback
loop. Whether that shift of emphasis survives in models that resolve the
processes omitted here is an open question, and one that the present
framework is not equipped to settle.

\section{Conclusions}
\label{sec:conclusion}

Within a conservative five-box representation of the thermohaline
circulation with an Indo-Pacific gateway, the composition of the interbasin
return flow does not enter the equilibrium problem at all: adding the
steady-state budgets of the gateway reservoirs cancels every internal
exchange, leaving the salt delivered to the Atlantic fixed by the net
Atlantic freshwater export, so that the warm-route fraction, the ITF
strength, the retroflection, and the apportionment of that export among
reservoirs are collectively immaterial to where the folds lie. Replacing
the discrete gateway transit lag by a normalized memory kernel leaves those
equilibria untouched, because the kernel integrates to unity, but confers
on the system a route of influence that the steady states deny it: the
dispersion of transit times sets a closed-form threshold for oscillatory
instability, broad memory is markedly more stabilizing than a discrete lag,
and because the instantaneous feedback weakens to nothing as the fold is
approached, that instability is encountered before the saddle-node under
every kernel sharper than a decaying exponential. The two findings are
complementary rather than merely coexistent, since the first closes the
equilibrium channel through which interbasin exchange has conventionally
been supposed to act and the second identifies the transient channel that
remains open. Whether this displacement of influence from the steady states
to the transients survives in models that resolve the eddy and gyre
pathways a well-mixed box formulation omits is the question that would
determine how far the present conclusions travel.

\section*{Declaration of competing interest}
The authors declare that they have no known competing financial interests or
personal relationships that could have appeared to influence the work
reported in this article.

\section*{Declaration of generative AI use}
During the preparation of this work, the authors used Claude Sonnet 5 solely for the purposes of English grammar, vocabulary refinement, and improving the overall readability of the manuscript. The authors maintain full responsibility for the conceptualization, model development, mathematical derivations, computational execution, and the analysis and interpretation presented in this study.

\section*{Code and data availability}
This study uses no external data. The analysis code is available at
\url{https://github.com/sandyherho/amoc_gateway_reduced}, where executing
\texttt{scripts/run\_all.py} regenerates every result reported here. All
generated outputs, comprising the figures in vector and raster form, the
numerical content of every figure panel as comma-separated value (CSV)
tables, and
the plain-text computation reports, are archived on the Open Science
Framework (OSF) at \url{https://doi.org/10.17605/OSF.IO/T6H4M}. Both the code
and the archive are released under the MIT License.

\section*{Funding}
This study was supported by the Faculty of Earth Sciences and Technology (FITB),
Bandung Institute of Technology (ITB), through the PPMI Research Program 2026
under Project ID FITB.PPMI-1-19-2026.

\section*{Acknowledgements}
We thank the developers and maintainers of NumPy, SciPy, and Matplotlib for
the open-source scientific computing infrastructure on which this work
depends.

\end{document}